\documentclass[aps,prl,twocolumn,showpacs,superscriptaddress]{revtex4}  % for review and submission
\usepackage{graphicx}  % needed for figures
\usepackage{dcolumn}   % needed for some tables
\usepackage{bm}        % for math
\usepackage{amssymb}   % for math
\usepackage[dvipdfm]{hyperref}
\hypersetup{
    colorlinks=true,
    linkcolor=blue,
    citecolor=blue,
    urlcolor=blue,
    pdftitle={Dissociation products and structures of solid H$_2$S at strong compression},
    bookmarks=true,
}

\begin{document}

% the following line is for submission, including submission to the arXiv!!
%\hspace{5.2in} \mbox{Fermilab-Pub-04/xxx-E}

\title{Dissociation products and structures of solid H$_2$S at strong compression}

\author{Yinwei Li} \affiliation{School of Physics and Electronic Engineering, Jiangsu Normal University, Xuzhou 221116, China}
\author{Lin Wang} \affiliation{Center for High Pressure Science and Technology Advanced Research, Shanghai, 201203, China} \affiliation{State Key Laboratory of Superhard Materials, Jilin University, Changchun 130012, China}
\author{Hanyu Liu} \affiliation{State Key Laboratory of Superhard Materials, Jilin University, Changchun 130012, China} \affiliation{Geophysical Laboratory, Carnegie Institution of Washington, Washington D.C. 20015, USA}
\author{Yunwei Zhang} \affiliation{State Key Laboratory of Superhard Materials, Jilin University, Changchun 130012, China}
\author{Jian Hao} \affiliation{School of Physics and Electronic Engineering, Jiangsu Normal University, Xuzhou 221116, China}
\author{Chris J.\ Pickard} \affiliation{Department of Materials Science $\&$ Metallurgy,
University of Cambridge, 27 Charles Babbage Road, Cambridge CB3 0FS, United Kingdom}
\author{Joseph R.\ Nelson} \affiliation{Theory of Condensed Matter Group, Cavendish Laboratory, J J Thomson Avenue, Cambridge CB3 0HE, United Kingdom}
\author{Richard J.\ Needs} \affiliation{Theory of Condensed Matter Group, Cavendish Laboratory, J J Thomson Avenue, Cambridge CB3 0HE, United Kingdom}
\author{Wentao Li} \affiliation{Center for High Pressure Science and Technology Advanced Research, Shanghai, 201203, China}
\author{Yanwei Huang } \affiliation{Center for High Pressure Science and Technology Advanced Research, Shanghai, 201203, China}
\author{Ion Errea} \affiliation{Donostia International Physics Center (DIPC), Manuel de Lardizabal pasealekua 4, 20018 Donostia-San Sebasti$\acute{a}$n, Basque Country, Spain} \affiliation{Fisika Aplikatua 1 Saila, EUITI Bilbao, University of the Basque Country (UPV/EHU), Rafael Moreno "Pitxitxi" Pasealekua 3, 48013 Bilbao, Basque Country, Spain}
\author{Matteo Calandra} \affiliation{IMPMC, UMR CNRS 7590, Sorbonne Universit$\acute{e}$s - UPMC Univ. Paris 06, MNHN, IRD, 4 Place Jussieu, F-75005 Paris, France}
\author{Francesco Mauri} \affiliation{IMPMC, UMR CNRS 7590, Sorbonne Universit$\acute{e}$s - UPMC Univ. Paris 06, MNHN, IRD, 4 Place Jussieu, F-75005 Paris, France}
\author{Yanming Ma}\email{mym@jlu.edu.cn} \affiliation{State Key Laboratory of Superhard Materials, Jilin University, Changchun 130012, China}

\begin{abstract}
  Hydrogen sulfides have recently received a great deal of interest
  due to the record high superconducting temperatures of up to 203 K
  observed on strong compression of dihydrogen sulfide (H$_2$S).  A
  joint theoretical and experimental study is presented in which
  decomposition products and structures of compressed H$_2$S are characterized, and their superconducting properties are calculated. In addition to the experimentally known H$_2$S and H$_3$S phases, our first-principles structure searches have identified several energetically competitive stoichiometries that have not been reported previously; H$_2$S$_3$, H$_3$S$_2$, and H$_4$S$_3$. In particular, H$_4$S$_3$ is predicted to be thermodynamically stable within a large pressure range of 25--113 GPa.  High-pressure room-temperature X-ray diffraction measurements confirm the presence of H$_3$S and H$_4$S$_3$ through decomposition of H$_2$S that emerge at 27 GPa and coexist with residual H$_2$S, at least up to the highest pressure studied in our experiments of 140 GPa. Electron-phonon coupling calculations show that H$_4$S$_3$ has a small \emph{T}$_c$ of below 2 K, and that H$_2$S is mainly responsible for the observed superconductivity of samples prepared at low temperature ($<$100K).
\end{abstract}

\maketitle

%\section{\label{sec:level1}First-level heading}
% sections are not used for PRL papers
% main text

Superconductivity with a transition temperature \emph{T}$_c$ of up to 203 K
was observed recently in solid H$_2$S at Megabar pressures, which
is the highest record among all known superconductors
\cite{drozdov_2015}.  Ashcroft suggested that metallic hydrogen would
be a superconductor at high pressures with a \emph{T}$_c$ around room
temperature \cite{ashcroft_1968}, and subsequently predicted that
hydrogen-rich metallic compounds might also be superconducting at high
pressures \cite{ashcroft_2004}.  Early theoretical studies focussed on
high-pressure silicon and aluminum hydrides
\cite{silane_2006,AlH3_2007}, and a number of potential
high-temperature superconductors have now been proposed in compressed
hydrogen-rich compounds, with \emph{T}$_c$s estimated in the range 40--250 K (e.g., GaH$_3$ \cite{gallane_2011,GaH3_2014}, SnH$_4$ \cite{stannane_2010}, GeH$_4$ \cite{germane_2008}, NbH$_4$ \cite{NbH4_2014}, Si$_2$H$_6$ \cite{disilane_2010},
SiH$_4$(H$_2$)$_2$ \cite{silanehydride_2010}, CaH$_6$
\cite{calciumhydride_2012}, YH$_3$ \cite{YH3_2009}, YH$_4$ and YH$_6$
\cite{yttrium_2015}). They have not been realized in practice, in part
because of demanding experimental challenges.

The high pressure phase diagram of H$_2$S has been studied
extensively. H$_2$S is a sister molecule of H$_2$O, and is the only
known stable compound in the H-S system at ambient pressure.
High-pressure diamond anvil cell experiments led to the discovery of a
metallic phase at about 96 GPa
\cite{shimizu_1991,endo_1994,endo_1996,endo_1998,fujihisa_1998,fujihisa_2004,sakashita_1997,shimizu_1992,shimizu_1995,shimizu_1997,loveday_2000,22}. However, partial decomposition of H$_2$S and elemental sulfur was observed in Raman \cite{loveday_2000} and XRD studies \cite{fujihisa_2004} at room
temperature above 27 GPa.  H$_2$S had not been considered as a
candidate for superconductivity because it was believed to dissociate
into elemental sulfur and hydrogen under high pressures
\cite{23}. Recent first-principles structure searches predicted
energetically stable metallic structures of H$_2$S above 110 GPa
\cite{24} and excluded dissociation into its elements. An estimated
maximum \emph{T}$_c$ for metallic H$_2$S of 80 K at 160 GPa was predicted
\cite{24}. Motivated by this study, Drozdov \textit{et al.}\
\cite{drozdov_2015} performed high-pressure experiments on solid H$_2$S looking for superconductivity and found an astonishing T$_c$
of 203 K at 155 GPa \cite{drozdov_2015}. H$_2$S shows
complex superconducting behavior at high pressures with the emergence of two different superconducting states. Samples prepared at low temperature (100 K) have a \emph{T}$_c$ of $\sim$30 K at 110 GPa at the onset of
superconductivity, which increases rapidly to a maximum value of 150 K
at 200 GPa, while samples at room temperature or above show a maximum
\emph{T}$_c$ of 203 K at 155 GPa.

Strobel \textit{et al.}\ synthesized another H-S compound, H$_3$S, by compressing a mixture of H$_2$S and H$_2$ above 3.5 GPa \cite{25}. The superconducting \emph{T}$_c$ of H$_3$S at 200 GPa was recently predicted to be as high as 191--204 K \cite{26}, with H$_3$S in a cubic Im$\bar{3}$m structure, which is known already in H$_3$O at terapascal pressures \cite{pickard_terapascal} through decomposition of H$_2$O.  The agreement between experimental \cite{drozdov_2015} and theoretical
values of \emph{T}$_c$ \cite{26} led to the proposal by Drozdov \textit{et
  al.}\ that H$_3$S could be formed by decomposition of H$_2$S and
might be responsible for the observed superconducting state at 203 K
\cite{drozdov_2015}. This proposal was supported by
first-principles density-functional-theory (DFT) studies which
suggested that it is thermodynamically favorable for H$_2$S to
decompose into H$_3$S + S \cite{27,28,29} at pressures above 43 GPa
\cite{28}. Apparently, there is an urgent need to characterize the decomposition products of compressed H$_2$S in an effort to build an understanding of the complex superconducting behavior exhibited by the H-S system.

Here we present a joint theoretical and experimental study of
compressed H$_2$S which clarifies the possible decomposition products
and their structures. First-principles DFT structure searches were
used to predict several new stoichiometries (H$_2$S$_3$, H$_3$S$_2$, and
H$_4$S$_3$) and a new structure in H$_3$S not reported previously. Room-temperature high pressure X-ray diffraction (XRD) experiments demonstrate that above 27 GPa, H$_2$S partially decomposes into S + H$_3$S + H$_4$S$_3$. H$_4$S$_3$ emerges as the major component at around 66 GPa and coexists with a small fraction of H$_3$S and residual H$_2$S, at least up to the highest pressure studied experimentally of 140 GPa.

Extensive structure searches over 44 H-S stoichiometries at 25, 50,
100 and 150 GPa were performed using the CALYPSO \cite{30, 31} and
AIRSS \cite{silane_2006, 33} methods, which have been successfully
used to investigate structures of materials at high pressures
\cite{calciumhydride_2012,silane_2006,34,35,36,37,38,39,40}. The
underlying structural relaxations were performed using the Vienna
\textit{ab initio} simulation package (VASP) \cite{41} for CALYPSO and
the CASTEP plane-wave code \cite{42} for AIRSS. Electron-phonon coupling (EPC) calculations were performed with density functional perturbation theory using the Quantum-ESPRESSO package\cite{pwscf}. XRD data were
collected at the 15U1 beamline at the Shanghai Synchrotron Radiation
Facility (SSRF) with a monochromatic beam of wavelength 0.6199
{\AA}. The diffraction patterns were integrated with the FIT2D
computer code \cite{43} and fitted by Rietveld profile matching using
the GSAS+EXPGUI programs \cite{44, 45}. More information about the
calculations and experiments is provided in the Supplemental Material
\cite{46}.

\begin{figure}
  \begin{center}
  \includegraphics[width=1.0\columnwidth]{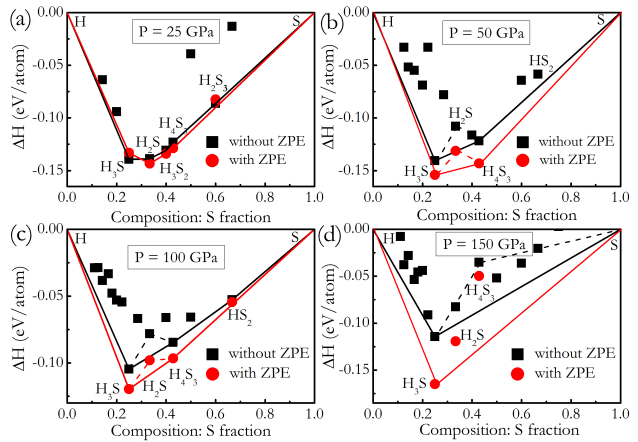}
  \caption{\label{fig:1} (Color online) Results from structure
    searching at 25 (a), 50 (b), 100 (c) and 150 GPa (d). Convex hulls
    are shown as continuous lines, with (red) and without (black) the
    inclusion of zero point vibrational enthalpy (ZPE).}
  \label{fig:Figure_1}
  \end{center}
\end{figure}

Figure \ref{fig:Figure_1} shows convex hull diagrams at 25, 50, 100
and 150 GPa which summarize the results of the structure searches. The
effects of including quantum zero-point vibrational motion are
significant and they tend to increase with pressure.  Our results
suggest that up to five H-S compounds lie on the convex hull at some
pressures and are therefore thermodynamically stable.  Besides the
experimentally known H$_2$S and H$_3$S compounds, we predict three
additional stable compounds: H$_4$S$_3$, H$_3$S$_2$ and H$_2$S$_3$. Note that H$_2$S is theoretically found to be stable only below 25 GPa,
while H$_3$S appears at all pressures considered. Enthalpy
calculations show that H$_3$S becomes energetically more stable than
H$_2$S+1/2H$_2$ at around 6 GPa. The newly predicted H$_3$S$_2$ and
H$_2$S$_3$ phases have very narrow pressure ranges of stability and
are unstable above 34 and 25 GPa, respectively (Supplemental Material,
Fig.\ S1 \cite{46}). We have therefore omitted further discussion of
these compounds.  The corresponding crystallographic parameters and
phonon dispersion curves are provided in the Supplemental Material
\cite{46}.

For H$_3$S, besides the P1, Cccm, R3m and Im$\bar{3}$m structures of
earlier studies \cite{26}, our searches predict a monoclinic C2/c
structure (4 f.u/cell) that was not reported earlier. The C2/c
structure consists of weakly bonded H$_2$S and H$_2$ molecules
(Supplemental Material, Fig.\ S7b \cite{46}), and is calculated to be
more stable at pressures of 2--112 GPa than the Cccm structure
proposed previously \cite{26} (Supplemental Material, Fig.\ S7a
\cite{46}). Static-lattice enthalpy calculations give a
zero-temperature phase sequence for H$_3$S of P1 $\rightarrow$ C2/c (2
GPa) $\rightarrow$ R3m (112 GPa) $\rightarrow$ Im$\bar{3}$m (180 GPa).

\begin{figure}
  \begin{center}
  \includegraphics[width=1.0\columnwidth]{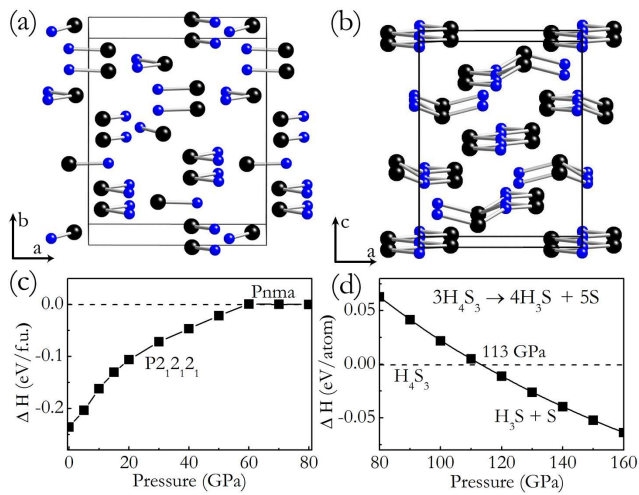}
  \caption{\label{fig:2} (Color online) Energetically
    favorable structures of ${\rm H_4S_3}$ with space groups
    $P2_12_12_1$ (a) and Pnma (b). Small and large spheres represent H and S atoms, respectively. (c) Calculated enthalpy curves
    of the $P2_12_12_1$ structure with respect to the Pnma
    structure for ${\rm H_4S_3}$ as a function of pressure. (d)
    Decomposition enthalpy curves of ${\rm H_4S_3}$ into (H$_3$S+S) as a
    function of pressure. \label{fig:Figure_2}}
  \end{center}
\end{figure}

H$_4$S$_3$ adopts an orthorhombic $P2_12_12_1$ (4 f.u./cell)
structure that consists of weakly bonded HS and H$_2$S molecules at 25 GPa
(Fig.~\ref{fig:Figure_2}a). The H-S bond lengths within the
HS and H$_2$S molecules are 1.354 {\AA} and 1.387-1.391 {\AA},
respectively, which are significantly shorter than the H-S separation
of 1.913-1.932 {\AA} between molecules. With increasing pressure, the
neighboring molecules bond with each other forming planar H-S-H-S
zigzag chains and puckered H-S-H-S chains,
respectively. $P2_12_12_1$ transforms to a Pnma structure at 60
GPa (Fig.~\ref{fig:Figure_2}c). The convex hull data suggests two
synthesis routes for H$_4$S$_3$: (i) decomposition of
8H$_2$S$\rightarrow$S + 4H$_3$S + H$_4$S$_3$ above 25 GPa; (ii)
reaction of 4H$_3$S + 5S $\rightarrow$ 3H$_4$S$_3$ in the pressure
range of 25--113 GPa. Theoretically, it is found that H$_4$S$_3$
decomposes into H$_3$S + S above 113 GPa (Fig.~\ref{fig:Figure_1}d and
Fig.~\ref{fig:Figure_2}d).

\begin{figure}
  \begin{center}
  \includegraphics[width=1.0\columnwidth]{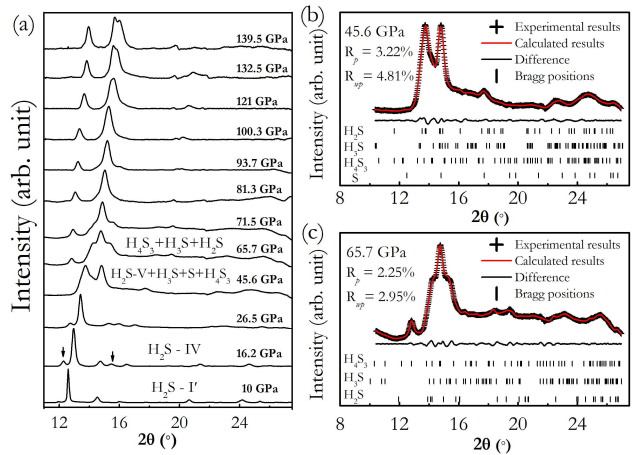}
  \caption{\label{fig:3} (Color online) (a) XRD patterns of H$_2$S
    collected at various pressures and at room temperature with an
    incident wavelength of 0.6199 {\AA}. [(b) and (c)] Rietveld
    refinements of XRD profiles at 45.6 GPa based on the compositions
    Pc-H$_2$S, I4$_1$/acd-S, C2/c-H$_3$S and
    $P2_12_12_1$-H$_4$S$_3$ with phase fractions of 1147:85:31:1,
    and at 65.7 GPa based on compositions of Pnma-H$_4$S$_3$,
    C2/c-H$_3$S and Pc-H$_2$S with phase fractions of 5:3.4:1,
    respectively. The cross symbols and red solid lines represent
    observed and fitted patterns, respectively. The solid lines at the
    bottom are the difference between the observed and fitted
    patterns. Vertical bars under the pattern represent the calculated
    positions of reflections arising from the
    compositions.  \label{fig:Figure_3}}
  \end{center}
\end{figure}

Powder XRD patterns obtained on increasing pressure from 10 to 140 GPa
at room temperature are shown in Fig.~\ref{fig:Figure_3}a. The XRD
patterns collected at pressures up to 46 GPa are in excellent
agreement with previous data
\cite{endo_1994,endo_1996,endo_1998,fujihisa_1998,fujihisa_2004}, and
the successive transitions of phase I$^{\prime}$ $\rightarrow$ phase
IV $\rightarrow$ phase V are well reproduced. The XRD data at 10 GPa
correspond to phase I$^{\prime}$. Phase IV with additional peaks at
around 12$^{\circ}$ and 15$^{\circ}$ (shown by arrows in
Fig.~\ref{fig:Figure_3}a) was observed at 16 GPa.  The diffraction
peaks of phase IV weaken at pressures above 27 GPa, and a new
diffraction profile observed at 46 GPa is responsible for phase V. The
XRD data show that the IV$\rightarrow$V transition begins above 27
GPa, in excellent agreement with previous results
\cite{endo_1996,fujihisa_2004}.

Previous high-pressure Raman \cite{loveday_2000} and XRD
\cite{fujihisa_2004} studies have claimed that decomposition of H$_2$S
occurs at room temperature above 27 GPa.  Indeed, we found that H$_2$S
partially decomposes in phase V. Unfortunately, we have not found it
possible to resolve the decomposition products and their crystal
structures from our current XRD data.  Therefore we use the predicted
structures and convex hull data to help in analysing the experimental
data. At 50 GPa, our calculations suggest an energetically allowed
dissociation path of 8H$_2$S$\rightarrow$S + 4H$_3$S + H$_4$S$_3$
(Fig.\ 1b).  The XRD profile at 46 GPa was therefore fitted to a
mixture of H$_2$S + S + H$_3$S + H$_4$S$_3$ by performing Rietveld
refinements using the most energetically stable
structures. Remarkably, we found that the XRD profile can be well
indexed by a mixture of Pc-structured H$_2$S, I4$_1$/acd-structured S,
C2/c-structured H$_3$S and $P2_12_12_1$-structured H$_4$S$_3$,
with phase fractions (ratios of numbers of unit cells) of
1147:85:31:1, as shown in Fig.~\ref{fig:Figure_3}b.  The existence of a large proportion of H$_2$S demonstrates the partial decomposition. We also
attempted other Rietveld refinements fitting the XRD patterns to pure
H$_2$S, H$_2$S + S + H$_3$S or S + 4H$_3$S + H$_4$S$_3$, but all of
these fits gave poorer results (Supplemental Material, Fig.\ S10
[46]).  The calculated decomposition pressure (30 GPa) for 8H$_2$S
$\rightarrow$ S + 4H$_3$S + H$_4$S$_3$ (Supplemental Material, Fig.\
S13 \cite{46}) is in excellent agreement with the value of 27 GPa
observed in experiment \cite{fujihisa_2004}.

\begin{figure}
  \begin{center}
  \includegraphics[width=1.0\columnwidth]{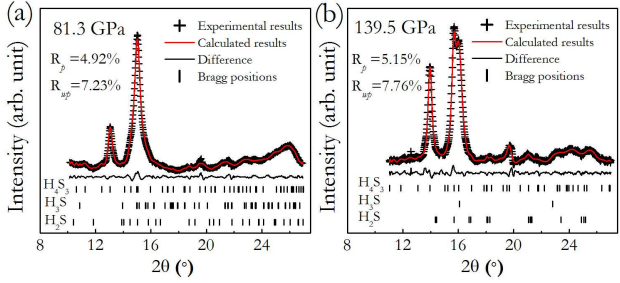}
  \caption{\label{fig:4} (Color online) Rietveld refinements of XRD
    profiles at 81.3 GPa based on Pnma-H$_4$S$_3$ + C2/c-H$_3$S +
    Pmc2$_1$-H$_2$S with phase fractions of 43:6:1 (a) and at 139.5
    GPa based on Pnma-H$_4$S$_3$ + R3m-H$_3$S + P-1-H$_2$S with phase
    fractions of 56:7:1 (c). \label{fig:Figure_4}}
  \end{center}
\end{figure}

With increasing pressure, more H$_2$S decomposes and its contribution
to the XRD signal is reduced. The XRD pattern collected at 66 GPa
shows entirely different features to that at 46 GPa. Rietveld
refinement shows that a mixture of Pnma-structured H$_4$S$_3$,
C2/c-structured H$_3$S, and Pc-structured H$_2$S with phase fractions
5:3.4:1 gives the best fit to the experimental data
(Fig.~\ref{fig:Figure_3}c). The disappearance of elemental S and the
reduction in the ratio of H$_3$S are understandable since a reaction
of 4H$_3$S + 5S $\rightarrow$ 3H$_4$S$_3$ takes place as inferred from
our convex hull calculations (Fig.~\ref{fig:Figure_1}b and
\ref{fig:Figure_1}c). The two shoulders on the main peak at
15$^{\circ}$ originate primarily from H$_3$S and H$_2$S. These
shoulders weaken when the pressure is increased to 82 GPa and the XRD
pattern can then be well indexed by Pnma-structured H$_4$S$_3$,
C2/c-structured H$_3$S and Pmc2$_1$-structured H$_2$S with a major
contribution (86\%) from H$_4$S$_3$ (Fig.~\ref{fig:Figure_4}a). Our
results demonstrate that H$_4$S$_3$ coexists with H$_3$S and H$_2$S at
least up to 140 GPa, the highest pressure studied experimentally
(Fig.~\ref{fig:Figure_4}b). At this pressure, a refinement based on
Pnma-structured H$_4$S$_3$ and C2/c-structured H$_3$S leads to poorer
fits with higher R\emph{$_{p}$} and R\emph{$_{wp}$} values
(Supplemental Material, Fig.\ S12 \cite{46}), which supports the
existence of residual H$_2$S (about 1.6\%).

\begin{figure}
  \begin{center}
  \includegraphics[width=1.0\columnwidth]{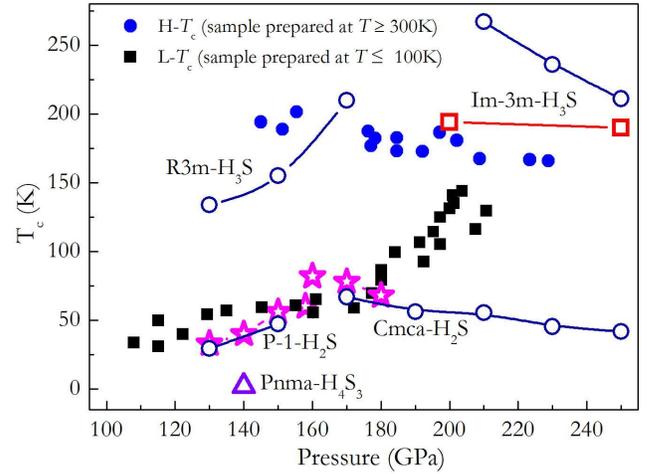}
  \caption{\label{fig:5} (Color online) Superconducting transition
    temperatures (\emph{T}$_c$) calculated for various H-S compounds and
    experimental values for compressed H$_2$S \cite{drozdov_2015}. Solid squares and circles show experimental data from Fig.\ 1(b) and Fig.\ 2 (b) of Ref.\ \onlinecite{drozdov_2015}, respectively, where different runs are represented by the same symbols. The open stars, squares and
    circles show calculated data from Ref.\ \onlinecite{24}, Ref.\
    \onlinecite{29} and Ref.\ \onlinecite{48}, respectively. Open
    triangles denote calculated \emph{T}$_c$ for Pnma-H$_4$S$_3$ in the
    present study. Note that the \emph{T}$_c$ values (open squares) for
    Im$\bar{3}$m-H$_3$S taken from Ref.\ \onlinecite{29} include
    anharmonic effects, while all other estimated \emph{T}$_c$s are
    calculated within the harmonic approximation.  \label{fig:Figure_5}}
  \end{center}
\end{figure}

H$_4$S$_3$ becomes metallic at 102 GPa (Supplemental Material, Fig.\
S14 \cite{46}). However, the calculated electron-phonon-coupling
parameter ($\lambda$ = 0.42) is very small at 140 GPa due to the low density of states at the Fermi level of 0.09 eV$^{-1}$/atom. As a result, the \emph{T}$_c$ estimated from the Allen and Dynes modified McMillan equation \cite{47} with $\mu^*$ of 0.16--0.13 is only 0.75--2.1 K at 140 GPa.

In Fig.~\ref{fig:Figure_5}, we compare the calculated \emph{T}$_c$ values for H$_4$S$_3$, H$_2$S and H$_3$S to experimental data measured in
compressed H$_2$S \cite{drozdov_2015}, where \emph{T}$_c$ obtained for
samples prepared at low and high temperatures are denoted by L-\emph{T}$_c$
and H-\emph{T}$_c$, respectively. On the one hand, the observed L-\emph{T}$_c$ \cite{drozdov_2015} at pressures below 160 GPa can only be
quantitatively reproduced by H$_2$S, while the nature of the rapidly
increasing L-\emph{T}$_c$ above 160 GPa remains unclear because the
calculated \emph{T}$_c$ values of H$_3$S, H$_4$S$_3$, and H$_2$S are too
high, too low, and tending to decrease, respectively. On the other
hand, although the values of \emph{T}$_c$ for Im$\bar{3}$m structured H$_3$S calculated within the harmonic approximation are much higher than the observed H-\emph{T}$_c$ \cite{48}, the inclusion of anharmonic effects reproduces the H-\emph{T}$_c$ data above 180 GPa \cite{29} quite well. However, at low pressures around 150 GPa, \emph{T}$_c$ values \cite{48} estimated for R3m-H$_3$S within the harmonic approximation are $\sim$45 K lower than the observed H-\emph{T}$_c$. Meanwhile, the predicted \emph{T}$_c$ for R3m-H$_3$S increases with pressure, in stark contrast to the experimental observation of a decrease in H-\emph{T}$_c$. Apparently, further study is greatly needed to disclose the steep \emph{T}$_c$ increase of L-\emph{T}$_c$ above 160 GPa and the high H-\emph{T}$_c$ at around 150 GPa.

We find that kinetics plays an important role in determining the
experimentally observed H-S structures. Theory suggests that H$_2$S
and H$_4$S$_3$ decompose above 25 and 113 GPa (Supplemental Material
Fig.\ S13 and Fig.\ 2d), respectively. However, H$_2$S and H$_4$S$_3$
are observed to persist up to at least 140 GPa. Large kinetic barriers
must therefore play a major role in suppressing decomposition at high
pressures, as has been found in other materials \cite{49,50,51,52}.

In summary, through first-principles structure searching calculations, we predict three new stable H-S compounds with stoichiometries H$_3$S$_2$, H$_2$S$_3$, and H$_4$S$_3$ and a new C2/c structure of H$_3$S, enriching the phase diagram of H-S systems at high pressures. The formation of H$_4$S$_3$ and H$_3$S was confirmed by XRD experiments to occur through decomposition of compressed H$_2$S above 27 GPa resulting in the products S + H$_3$S + H$_4$S$_3$. H$_4$S$_3$ becomes a major component at around 66 GPa and is stable up to at least 140 GPa, with a small fraction of H$_3$S and residual H$_2$S. We have also examined potential superconductivity of metallic H$_4$S$_3$ via explicit calculations of electron-phonon coupling parameters and the superconducting \emph{T}$_c$. Our work suggests that kinetically protected H$_2$S in samples prepared at low temperature is responsible for the observed superconductivity below 160 GPa.

% Acknowledgement paragraph in English of Oct. 15, 2014 for APS journals
Y.\ L.\ and J.\ H.\ acknowledge funding from the National Natural
Science Foundation of China under Grant No.\ 11204111 and No.\
11404148, the Natural Science Foundation of Jiangsu province under
Grant No.\ BK20130223, and the PAPD of Jiangsu Higher Education
Institutions. Y.\ Z.\ and Y.\ M.\ acknowledge funding from the
National Natural Science Foundation of China under Grant Nos.\
11274136 and 11534003, the 2012 Changjiang Scholars Program of China. R.\ J.\ N.\ acknowledges financial support from the Engineering and Physical
Sciences Research Council (EPSRC) of the U.K.\ [EP/J017639/1]. Calculations were performed on the Cambridge High Performance Computing Service facility and the HECToR and Archer facilities of the U.K.'s national high-performance computing service (for which access was obtained via the UKCP consortium [EP/K013564/1]). J.\ R.\ N.\ acknowledges financial support from the Cambridge Commonwealth Trust. I.\ E.\ acknowledges financial support from the Spanish Ministry of Economy and Competitiveness
(FIS2013-48286-C2-2-P). M.\ C.\ acknowledges support from the Graphene
Flagship and Agence nationale de la recherche (ANR), Grant No.\
ANR-13-IS10-0003-01. Work at Carnegie was partially supported by
EFree, an Energy Frontier Research Center funded by the DOE, Office of
Science, Basic Energy Sciences under Award No.\ DE-SC-0001057 (salary
support for H.\ L.). The infrastructure and facilities used at
Carnegie were supported by NNSA Grant No.\ DE-NA-00006, CDAC.

\end{document}